\documentclass[journal]{IEEEtran}
\usepackage{cite}
\usepackage{graphicx, caption, sidecap}
\usepackage[caption=false,font=footnotesize]{subfig}
\usepackage{fixltx2e}

\usepackage{amsmath,amssymb}
\newcommand{\Mod}[1]{\ (\text{mod}\ #1)}

\begin{document}

\title{Automatic Resonance Alignment of High-Order Microring Filters}
\author{Jason C. C. Mak, Wesley D. Sacher, Tianyuan Xue, Jared C. Mikkelsen, Zheng Yong, Joyce K. S. Poon ~\IEEEmembership{Member,~IEEE}%

\thanks{The authors are with the Department of Electrical and Computer Engineering, University of Toronto, Toronto, Ontario, Canada  (email: jcc.mak@mail.utoronto.ca; wesley.sacher@mail.utoronto.ca; jumpingjack.xue@mail.utoronto.ca; jared.mikkelsen@mail.utoronto.ca; zheng.yong@mail.utoronto.ca; joyce.poon@utoronto.ca).}% <-this % stops a space
}%\thanks{Manuscript received XXX, 2015}}

%% The paper headers
%\markboth{Journal of Quantum Electronics,~Vol.~XX, No.~XX, ~2015}%
%{Shell \MakeLowercase{\textit{et al.}}: Bare Demo of IEEEtran.cls for Journals}

\maketitle

\begin{abstract}

Automatic resonance alignment tuning is performed  in high-order series coupled microring filters using a feedback system. By inputting only a reference wavelength, a filter is tuned such that passband ripples are dramatically reduced compared to the initial detuned state and the passband becomes centered at the reference.  The method is tested on  $5^{\mathrm{th}}$ order  microring filters fabricated in a standard silicon photonics foundry process.  Repeatable tuning is demonstrated for filters on multiple dies from the wafer and for arbitrary reference wavelengths within the free spectral range of the microrings.
\end{abstract}


\IEEEpeerreviewmaketitle

%%%%%%%%%%%%%%%%%
%%%% INTRODUCTION %%%%%
%%%%%%%%%%%%%%%%%
\section{Introduction}
\IEEEPARstart{S}{\MakeLowercase{ilicon}}-on-insulator (SOI) has emerged in recent years as an attractive integrated photonics platform.  The availability of large wafer sizes, compatibility with complementary metal oxide semiconductor (CMOS) manufacturing, potential integration with CMOS electronics, and a high refractive index contrast opens the avenue for high volume production of densely integrated and compact microphotonic circuits  \cite{SorefJSTQE2006, Hochberg2010b, LimJSTQE2014, orcutt2008demonstration, kimerling2006electronic}.  However, a significant challenge for high index contrast silicon photonics is the management of fabrication variation owing to the sensitivity of the effective index to nanometer-scale dimensional variations \cite{PopovicCLEO2006}. Variability in device characteristics limits the achievable complexity of a system.  Thus, variation tolerant design, tunability, and feedback control are needed for the reliable performance of complex Si photonic circuits \cite{PadmarajuOE2012, Morichetti2014b, MikkelsenOE2014, mikkelsen2014dimensional, SuOFC2014, CoxOE2014, GrillandaOPT2014}.

\begin{figure}[t!]
\centering
\includegraphics[width = \columnwidth]{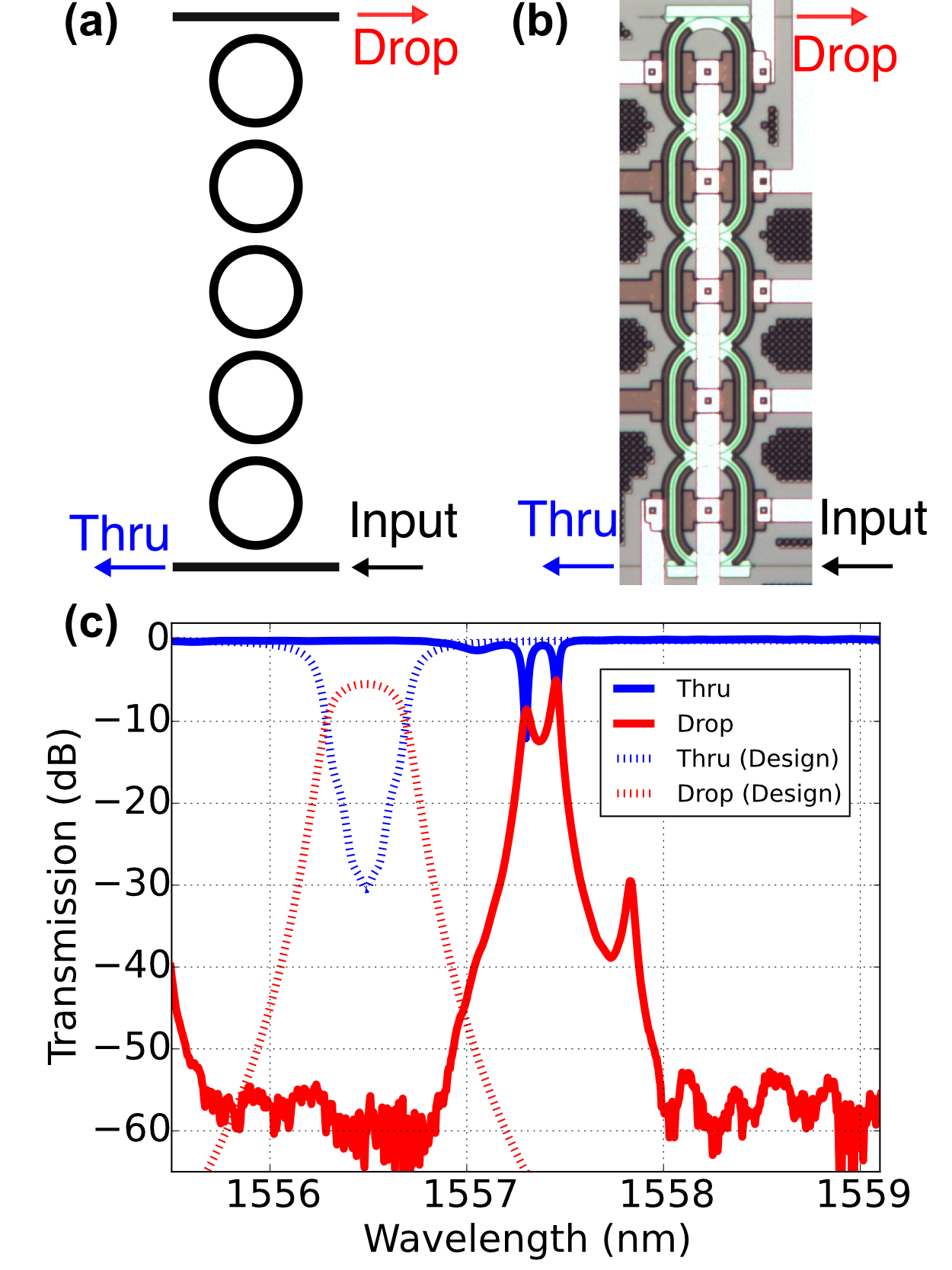}
\caption{(a) A schematic of a series coupled multi-ring optical filter. (b) A microscope image of a foundry fabricated 5-ring filter. Metallization for the thermal tuners is visible overlaying the microrings. (c) The measured drop and thru port spectra of the as-manufactured device in (b) is contrasted with the expected designed spectra (dots). Fabrication variation leads to ripples in the spectra and a detuned operation wavelength.}
\label{fig:foundry_variation}
\end{figure}

A class of photonic devices severely affected by fabrication variation are the high-order series coupled microring filters illustrated in Fig. \ref{fig:foundry_variation}(a). Fig. \ref{fig:foundry_variation}(b) shows an optical micrograph of a 5-ring filter fabricated at the A*STAR IME Si photonics foundry. Microring filters in the configuration as in Figs. \ref{fig:foundry_variation}(a) and (b) can be used as high-performance wavelength-division multiplexing (WDM) add-drop multiplexers with sharp filter roll-offs, flat-top passbands, and high out-of-band rejection \cite{Little2004, BarwiczJLT2006, Xia2007, Dong2010, OngPTL2013}. While the theoretical transfer function of the filter promises high performance WDM add-drop multiplexers suitable for dense integration, typical foundry as-manufactured devices cannot be used without tuning the ring resonances due to manufacturing variability.  Typical output spectra at the through (thru) and drop ports, corresponding to the device in Fig. \ref{fig:foundry_variation}(b), are shown in Fig. \ref{fig:foundry_variation}(c).  The measured spectra are vastly different from the design, with large ripples in the passband and noticeable detuning from the designed center wavelength.  
 
To recover the desired passband, the resonances of the microrings could be aligned  \cite{Dong2010, Lambert2012, Schrauwen2008}, using, for example, the thermo-optic effect to tune the refractive index \cite{popovic2007tunable, Amatya2007}. However, for sophisticated microring filters  to be suitable for  volume production, we need a resonance alignment method that can accommodate a wide variability in the initial detunings. One approach is to use a feedback system that automatically aligns the resonances of high-order microring filters implemented with electronic components. 

In this work, we demonstrate automatic resonance tuning of high-order microring filters fabricated at a Si photonic foundry using feedback. To our knowledge, this is the first demonstration of a system which automatically aligns the resonances of high-order microring filters.  The passband response is recovered reliably, over multiple instances of the device. Our method does not require an optical spectrum analyzer.  The position of the filter passband is defined by a reference wavelength, which is the optical input to the device. The automatic tuning system demonstrated here uses off-the-shelf components, though it can  be implemented as part of an embedded controller  packaged or integrated at the chip-scale level with the photonic chip. 
The approach used in this work can be extended to multi-ring filters of an arbitrary number of microrings, and possibly to other configurations of very large scale photonic integrated circuits.  

This paper is organized as follows.  In Section \ref{sect:theory}, we begin by examining the transfer function of the multi-ring filter to derive a controller for performing the resonance alignment, and verify its effectiveness on a model of the designed 5-ring filter. We then present the experimental procedure, setup, and results in Section \ref{sect:expt}.  The implications of the results are discussed in Section \ref{sect:discuss}.

%%%%%%%%%%%%%%
%%%% THEORY %%%%%
%%%%%%%%%%%%%%
\section{Theory}\label{sect:theory}

The resonance alignment procedure uses feedback, in which a fraction of the output from the ring device is measured and processed by a controller in order to determine successive tunings until  the desired resonance alignment is reached. A mathematical model of the device behavior is needed to formulate a controller algorithm. In Section \ref{sec:drop port phase detuning map}, we describe how the phase alignment of the microrings can be determined from the drop port power, and use this to formally state the resonance alignment problem for certain well chosen filters. In Section \ref{sec:optimization algorithm}, we describe controller algorithms for the resonance alignment, illustrating with an example 2-ring filter.

\subsection{Drop Port Phase Detuning Map}
\label{sec:drop port phase detuning map}

\begin{figure}[t]
\centering
\includegraphics[width = \columnwidth]{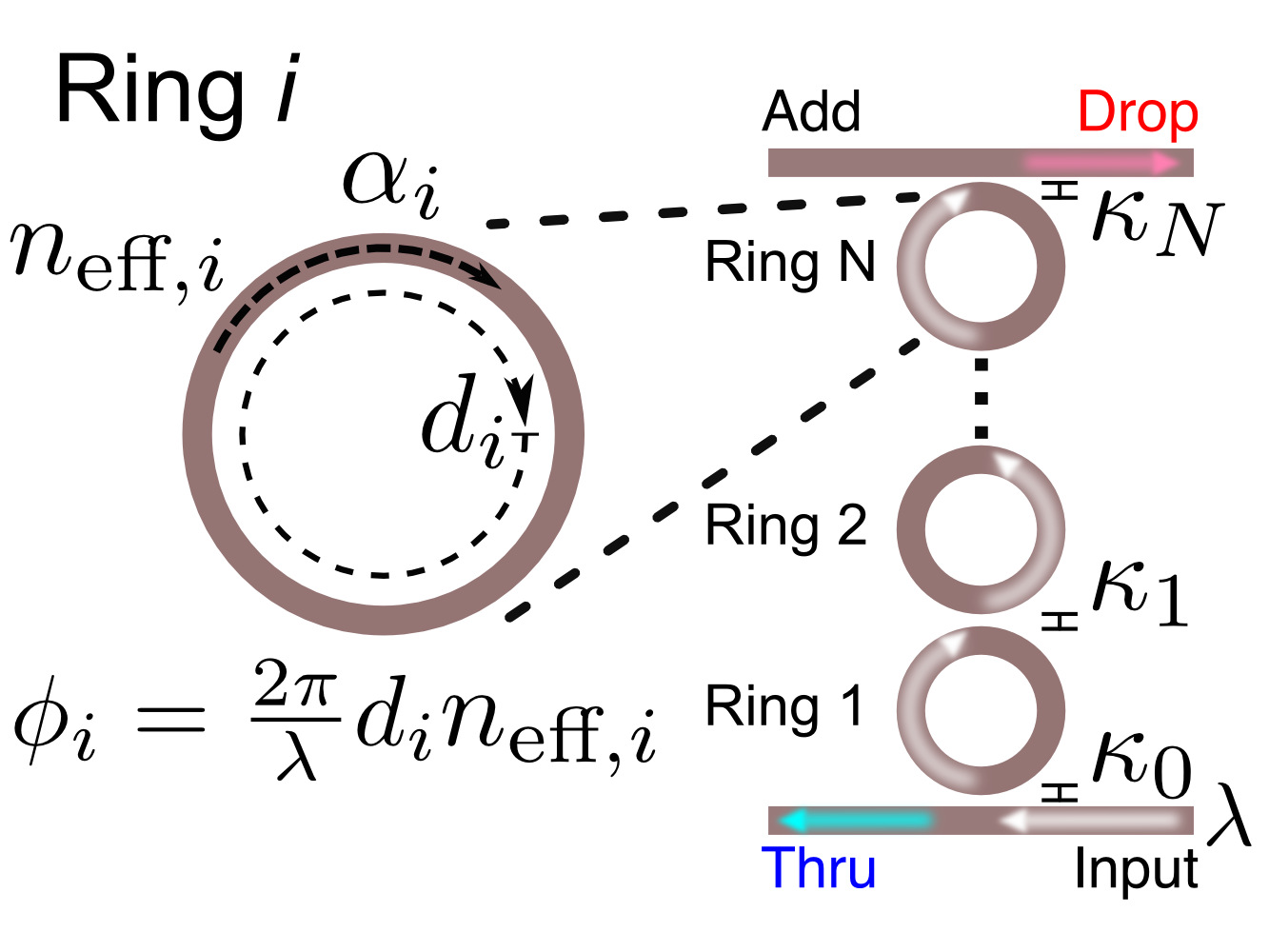}
\caption{Schematic of an $N$-ring filter, with quantities referred to in Eq. \ref{eq:drop port transfer}. The Drop port is specified for an odd number of rings (e.g., $N= 5$); for an even number of rings, the Add and Drop ports are swapped in the schematic. The terms  $\phi_i$, $n_{\textrm{eff},i}$, $\kappa_i$, and $\alpha_i$ have wavelength dependence. The cross-coupling coefficients between the rings, $\kappa_i$, are amplitude cross coupling coefficients. The shortest optical path to the Drop port is highlighted by white arrows on the rings.}
\label{fig:ring_schematic}
\end{figure}

\begin{figure}[t]
\centering
\includegraphics[width = \columnwidth]{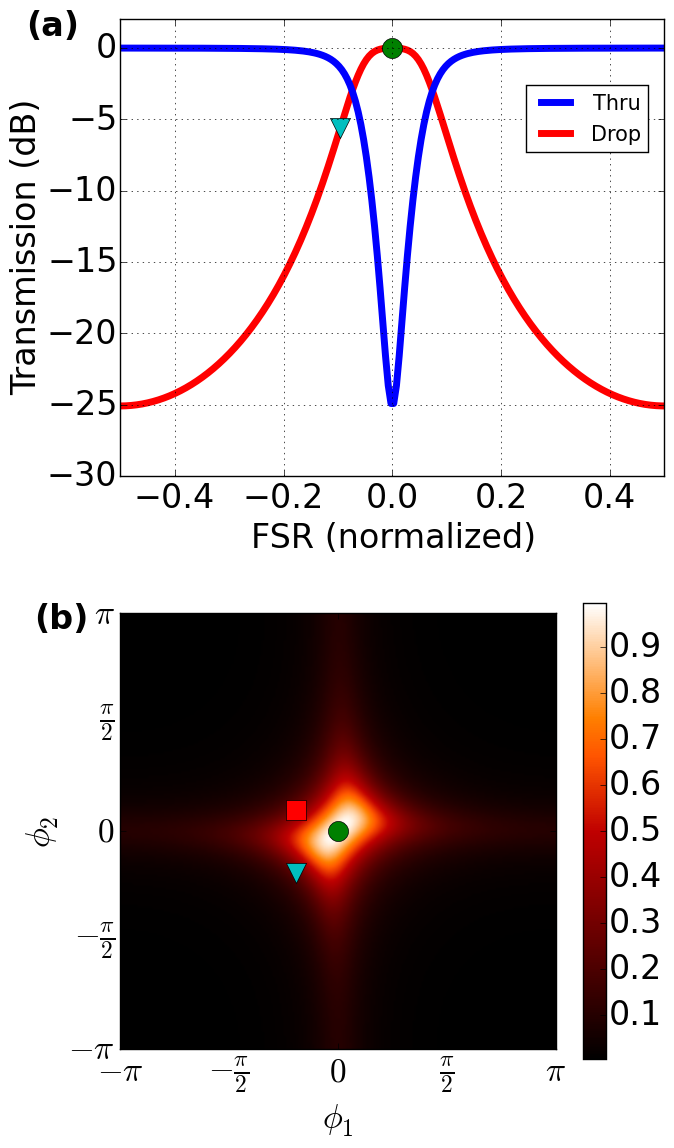}
\caption{(a) A 2-ring filter design with   $\kappa_0 = \kappa_2 = \sqrt{0.5}$, $\kappa_1 = \sqrt{0.1}$ and $\alpha = [1,1]$.  (b) The corresponding drop port power phase detuning map, $|T_D(\phi)|^2$. The green circle marks the transmission on resonance when the two rings are identical. A diagonal slice of $ |T_D(\phi)|^2$  corresponds to the drop port power in (a). The blue triangle marker corresponds to an input wavelength detuned from resonance for two identical rings.   The red marker shows an example with a random phase detuning.
}
\label{fig:detuning_map_example}
\end{figure}

\begin{figure*}%[t!] %use * environment to make it span across 2 pages?
\centering
\includegraphics[width = \textwidth]{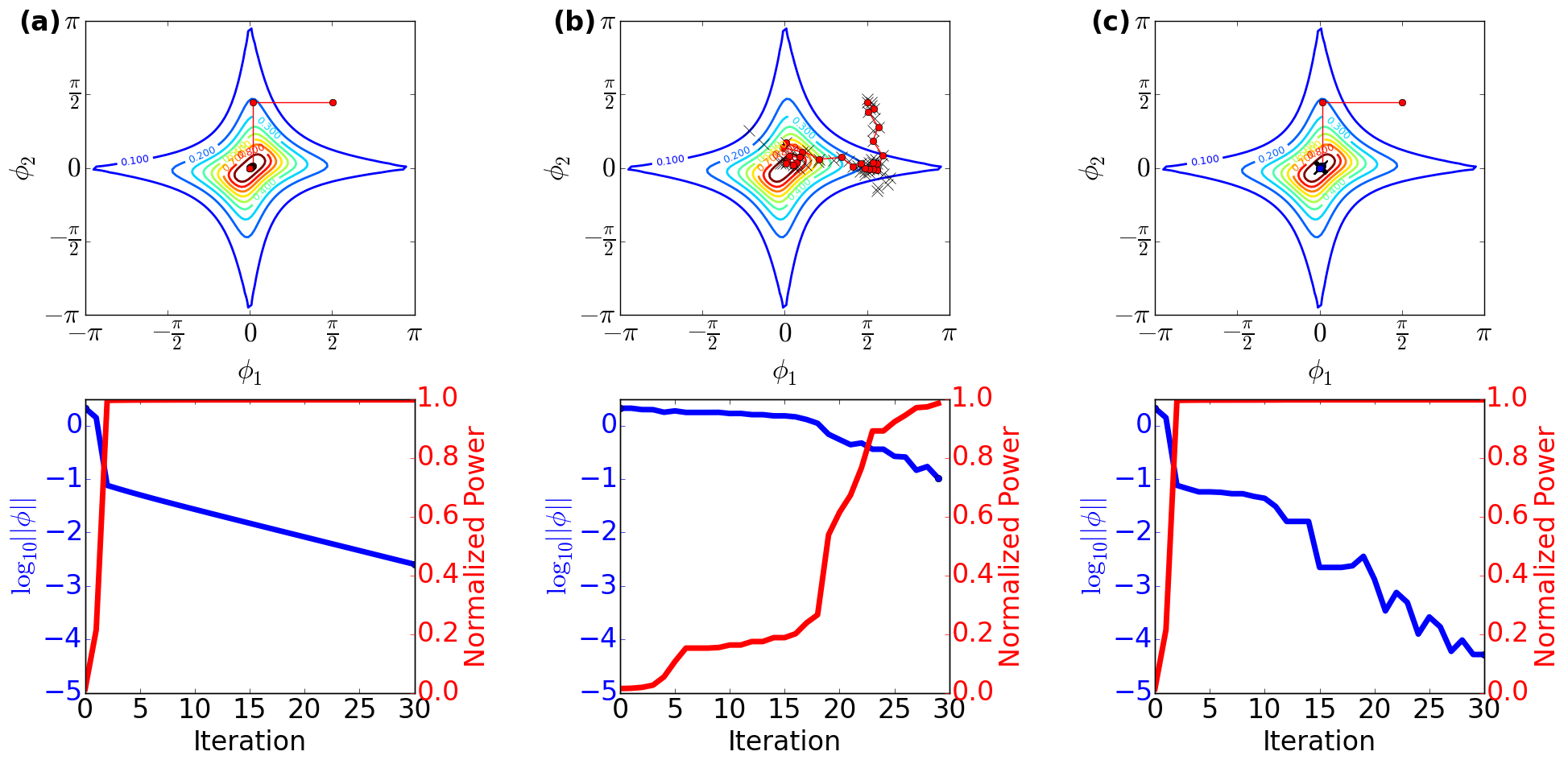}
\caption{Visualizations of the convergence (top row) and points on the detuning map (bottom row) for (a) coordinate decent, (b) Nelder-Mead, and (c) a combination of the two algorithms.}
\label{fig:simulation_2ring}
\end{figure*}

The  drop port transfer function of a $N$-ring filter illustrated in Fig. \ref{fig:ring_schematic} is derived in \cite{madsen1999optical, poon2004matrix} and is given by
\begin{equation}
\label{eq:drop port transfer}
T_D(\lambda) = \frac{\prod_{\ell=0}^{N}{-j\kappa_\ell}\sqrt{\prod_{m=1}^{N}{\alpha_m}}e^{-j\sum_{n=1}^{N}{\phi_n(\lambda)/2}}}{\xi_N(\phi(\lambda); \kappa,\alpha)}.
\end{equation}
In the above equation, light  within the $i$-th ring travels a path length of $d_i$ with a waveguide effective index of $n_{\textrm{eff},i}(\lambda)$, such that input light of wavelength $\lambda$ accumulates a round-trip phase-shift of $\phi_i(\lambda) = 2\pi n_{\textrm{eff},i}(\lambda) d_i/ \lambda$. The  field attenuation coefficient of the $i$-th ring is $\alpha_i$.   The coupling between adjacent rings is represented by the field cross-coupling coefficient $\kappa_i$ between the $i$-th and $(i-1)$-th ring. In general, $\alpha_i$ and $\kappa_i$ are wavelength dependent, but are typically approximated as constant. For brevity, we write in vector form the phase of the rings, $\phi = [\phi_1, \dots \phi_i \dots, \phi_N]$, the coupling coefficients, $\kappa = [\kappa_0, \dots, \kappa_{N}]$, and the attenuation constants, $\alpha = [\alpha_1, \dots, \alpha_N]$. As detailed in \cite{madsen1999optical}, assuming that all the rings are identical so that $\phi_i = \tilde\phi$ for all rings, the denominator $\xi_N$ is a polynomial of $z^{-1} = e^{-j\tilde\phi}$, and the transfer function is that of a digital all-pole filter and the roots of $\xi_N$ are the poles of the filter, with coefficients based on the $\kappa_i$. Thus, the flatness and the bandwidth of the drop port spectrum can be designed by choices of $\kappa$, while the round-trip path length sets the free spectral range (FSR) and centre wavelength.  

In Eq. \ref{eq:drop port transfer}, if we set $\lambda$ to be a constant  and  let $\phi_i$ be independent variables, then Eq. \ref{eq:drop port transfer} is a function of $N$ variables, which we write as $T_D(\phi)$.   $n_{\textrm{eff},i}$, $\kappa_i$, and $\alpha_i$ are constant at a fixed value of $\lambda$.  Assuming we can change the round-trip phase of the light in the rings, then at the fixed wavelength, the drop port transmission will vary with respect to the phase in each ring according to  $|T_D(\phi)|^2$, which we call the ``drop port power detuning map.''  $|T_D(\phi)|^2$ can be readily measured with a photodetector.  

 If the rings are identical, $\phi_i = \phi_\textrm{intrinsic}$, where $ \phi_\textrm{intrinsic}$ is the designed phase-shift.   An input wavelength that is on resonance would correspond to $\phi_i = 0 \Mod{2 \pi}$.  For illustration, we  consider the 2-ring filter design in Fig. \ref{fig:detuning_map_example}(a), where $\kappa_0 = \kappa_2 = \sqrt{0.5}$, $\kappa_1 = \sqrt{0.1}$ are approximated to be constant and the rings are lossless. The corresponding $|T_D(\phi)|^2$ is two dimensional and is shown in Fig. \ref{fig:detuning_map_example}(b). The drop port transmission can be determined from the plot for an arbitrary combination of $\phi_1$ and $\phi_2$.  If the two rings are identical, an on-resonance input  would be denoted by the green circular marker, and detuning from the resonance is shown by the blue triangle marker in Fig. \ref{fig:detuning_map_example}.  A diagonal slice of $|T_D(\phi)|^2$  corresponds to the computed drop port transmission spectrum in Fig. \ref{fig:detuning_map_example}(a). 

If the rings are not identical due to geometry variations that change  the effective index and ring dimensions, the phase-shift of each ring is
\begin{equation}
\label{eq:phase + variation}
\phi_i = \phi_{\textrm{intrinsic}} + \Delta\phi_{\textrm{intrinsic},i},
\end{equation}
where the $\Delta\phi_{\textrm{intrinsic}}$ is the phase deviation.  Thus, the problem of resonance alignment is to find the tuning terms, $\phi_{\textrm{tuning},i}$ to compensate for the variation, so that the net phase-shift on resonance for each ring is 
\begin{equation}
\label{eq:phase + variation}
\phi_i = \phi_{\textrm{intrinsic}} + \phi_{\textrm{tuning}, i} + \Delta\phi_{\textrm{intrinsic},i} = 0 \Mod {2\pi}.
\end{equation}
$|T_D(\phi)|^2$ is used to identify when the condition $\phi = 0 \Mod{2\pi}$ is satisfied for the rings. 

We further assume that the multi-ring filter has been suitably designed through the choice of coupling coefficients so the drop port transmission  is maximum at $\phi = 0 \Mod{2\pi}$. In the 2-ring filter example in Fig. \ref{fig:detuning_map_example}, the green circle marker is also at the maximum value within $[-\pi, \pi] \times [-\pi, \pi]$. For these designs, detunings caused by geometry variations would reduce the drop port transmission. Therefore, resonance alignment is to find the correct compensation phase tuning to maximize the drop port power. Visually, this corresponds to starting at an arbitrary, non-zero initial position on the drop port power detuning map (e.g., the red marker in  Fig. \ref{fig:detuning_map_example}(b)), and searching for the maximum point (i.e., the green marker in  Fig. \ref{fig:detuning_map_example}(b)). This is exactly the optimization problem
\begin{equation}
\label{eq:optimization problem}
\begin{aligned}
& \underset{\phi_\textrm{tuning}}{\text{maximize}}
& & |T_D(\phi_{\textrm{intrinsic}} + \Delta\phi_{\textrm{intrinsic}} +\phi_\textrm{tuning})|^2.
\end{aligned}
\end{equation}

Therefore, an optimization algorithm  can be thought of as the controller algorithm to move $\phi_\textrm{tuning}$ to the desired phase compensation values. An optimization algorithm  typically samples values of the objective function, $|T_D(\phi)|^2$, at various $\phi$ and iteratively moves toward the value of $\phi$ that  maximizes $|T_D(\phi)|^2$. Thus, the  feedback control system for the resonance alignment consists of an input, which is the sensor reading for $|T_D(\phi)|^2$, a controller, which is an optimization algorithm, and an output, which is the adjustment of $\phi$ of the device. While, in principle, any filter design with a transmission is maximum on resonance might be tuned in this manner, certain objective functions are more easily optimized. Roughly, $|T_D(\phi)|^2$ functions that are unimodal, with a well defined peak and few other surrounding local maxima, are more amenable to finding the optimal phase tuning without becoming trapped in a local maxima.  In practice, filters with smooth passbands have  $|T_D(\phi)|^2$  that are typically isolated and have few local maximum nearby, so  they are especially suitable for resonance alignment with this optimization approach.

\subsection{Optimization Algorithm Selection}
\label{sec:optimization algorithm}

In selecting the optimization algorithm, we need to consider the non-idealities of a real system, such as the noise of the detected signal and the discretization of $\phi$ and  $|T_D(\phi)|^2$ at the analog-to-digital and digital-to-analog converters for the microcontroller. Particularly, if the drop port transmission is low (i.e., when the rings are close to $\pi$ out of phase), noise will dominate the signal. Reducing noise through averaging creates a tradeoff between the operation speed and accuracy. Thus, an algorithm that is robust to noise and discretization is useful to reduce the number of evaluations and time required for the optimization.

We consider two optimization strategies, coordinate descent and the Nelder-Mead simplex algorithm, for the simplicity of implementation and to address the trade-off between speed and accuracy. Coordinate descent cycles through the coordinate directions, reducing the problem to a series of one dimensional optimization subproblems along each coordinate direction (i.e., a line-search), using the optimum for the current direction as the starting point  for the search in the next direction \cite{nocedal2006numerical}.  Coordinate descent by convention refers to minimization, but we use it for maximization by minimizing the negative of the objective function. The Nelder-Mead simplex algorithm \cite{nelder1965simplex} samples the objective function to enclose a generalized triangular volume (a simplex), and procedurally relocates the vertices of this volume to optimize the value of its centroid. The relocation involves reflection and scaling operations.  The simplex algorithm is a derivative free method, which, in general, is better suited to noisy or discrete optimization problems than gradient based methods \cite{nocedal2006numerical}. 

Fig. \ref{fig:simulation_2ring} illustrates the two methods applied to the 2-ring filter of Fig. \ref{fig:detuning_map_example}. For the calculations, we start from  $\phi = [1.58, 1.41]$.  The convergence of $||\phi|| \rightarrow 0$ and the normalized value of $|T_D(\phi)|^2$, which has a maximum of $0.996$, are plotted. The top row in Fig. \ref{fig:simulation_2ring} shows $\phi$ in red markers during the optimization process, and the bottom row shows the convergence. Although the effects of noise and finite resolution were not included, we observe the differences in performance of the two optimization methods.  Fig. \ref{fig:simulation_2ring}(a) shows the results for coordinate descent. The algorithm consecutively finds the maximum power by sweeping through the tuning of each ring, returning to the first ring after each cycle. Coordinate descent quickly finds the peak power but is slow to converge to the resonance.  Fig. \ref{fig:simulation_2ring}(b) shows the results for the Nelder-Mead  algorithm, and the locations of the vertices of the simplexes are marked with $\times$ in the top figure. $||\phi||$ converges more quickly compared to coordinate descent when it is close to the optimal value.  Therefore, we propose to use a combination of these two strategies for efficient convergence. In Fig. \ref{fig:simulation_2ring}(c), we initially perform 4 iterations of the coordinate descent before applying 26 iterations of the Nelder-Mead algorithm. For a comparable number of iterations, the combined strategy approaches the resonance the closest.

Some  differences exist between the simulation and the physical implementation. First,  in the simulation, the $\phi_i$ is directly set, but in experiments, we only have access to the voltage applied to the heaters, which is approximately proportional to the square root of $\phi_i$ (the proportionality is not exact due to dispersion). Second, the heaters may cause thermal cross-talk between adjacent rings.  However,  the Nelder-Mead algorithm automatically accounts for these  transformations in the objective function of the optimization, and the algorithm still searches for the maximum $|T_D(\phi)|^2$. Therefore, the overall algorithm used in the implementation is as follows: First, we apply coordinate descent  to quickly approach regions where $|T_D(\phi)|^2$ is higher to improve the signal-to-noise ratio. In the coordinate descent, the method used for the line-search subproblem is a brute force sweep.  Then, the Nelder-Mead algorithm is  applied to more quickly reach the final  tuning settings that maximize $|T_D(\phi)|^2$. 

%%%%%%%%%%%%%%
%%%% Experiment %%%%%
%%%%%%%%%%%%%%
\section{Experiment}\label{sect:expt}
\begin{figure}[t!]
\centering
\includegraphics[width = \columnwidth]{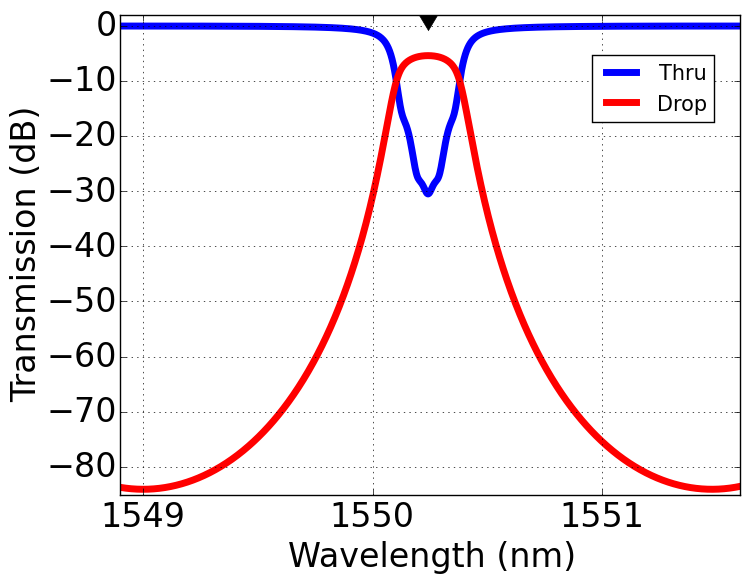}
\caption{The calculated 5-ring filter design used for the experiments, with constant cross coupling coefficients of $\kappa_0 = \kappa_5 = \sqrt{0.5}$, $\kappa_1 = \kappa_4 = \sqrt{0.07}$, $\kappa_2 = \kappa_3 = \sqrt{0.04}$. The center wavelength is  marked with a black notch at 1550.24 nm.  The waveguide loss was taken be 20 dB/cm. $\Delta \lambda_{\mathrm{1 dB}}$ is 0.174 nm, and $\Delta \lambda_{\mathrm{3 dB}}$ is 0.248 nm.}
\label{fig:filter_design}
\end{figure}
\begin{figure}[!htbp]
\centering
\includegraphics[width = \columnwidth]{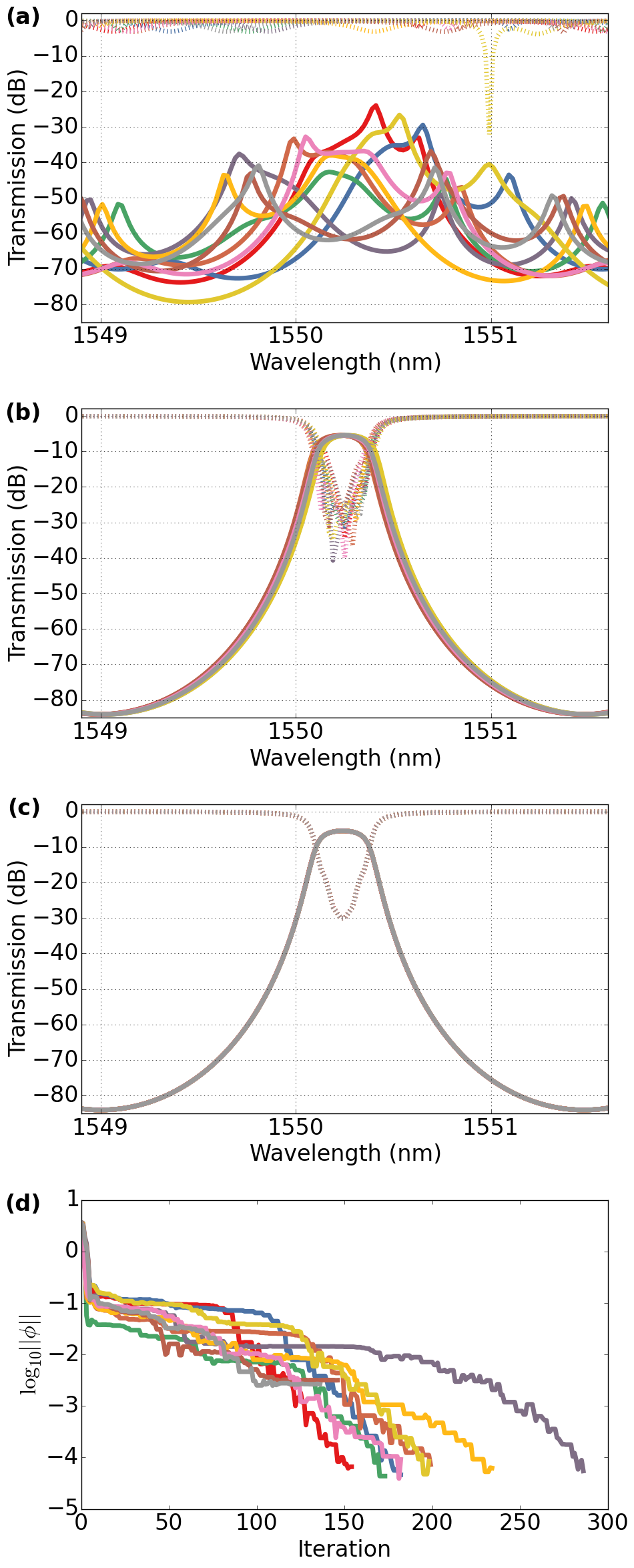}
\caption{Numerical experiment to show the convergence of 10 random initial detunings for a 5-ring filter. Spectra of the (a) the initially detuned microrings, (b) after coordinate descent, (c) and after Nelder-Mead (the spectra are overlapped atop each other).   (d) The convergence plot of the ring round trip phases $\phi$.}
\label{fig:simulation_5ring_randominitial}
\end{figure}

\subsection{Device Design}
\label{sect:device design}

We used a 5-ring filter to demonstrate the automatic resonance alignment system.  The designed coupling coefficients were $\kappa_0 = \kappa_5 = \sqrt{0.5}$, $\kappa_1 = \kappa_4 = \sqrt{0.07}$, $\kappa_2 = \kappa_3 = \sqrt{0.04}$. The dispersion of $\kappa_i$ was neglected. The path length of each ring was taken to be $d_i = 263.5$ $\mathrm{\mu m}$ to match the fabricated device to be described below.  The ring waveguides were taken to be Si ribs with a 220 nm height, a 90 nm thick partially-etched slab, and a width of 500 nm surrounded by a SiO$_2$ cladding.  The effective and group indices of the waveguide near 1550 nm are 2.524 and 3.671, respectively.  
The designed filter drop port spectrum has a center wavelength of 1550.24 nm with an FSR of 2.47 nm (308 GHz). The waveguide loss, dominated by doping for thermal tuners, was estimated to be 20 dB/cm, resulting in an insertion loss of -5.36 dB. The 1 dB bandwidth, $\Delta \lambda_{\mathrm{1 dB}}$, is 0.174 nm (21.7 GHz), and 3 dB bandwidth $\Delta \lambda_{\mathrm{3 dB}}$ is 0.248 nm (30.9 GHz).  The designed transmission spectra are shown in Fig. \ref{fig:filter_design}.

%describe this in more detail
We tested the  two-step optimization strategy from \ref{sec:optimization algorithm} on this model of the 5-ring filter to ensure that $\phi = 0 \mod{2\pi}$ is the only attractive point.  Starting from 10,000 uniformly sampled initial detunings within $[-\pi, \pi]^5$, applying 10 iterations of coordinate descent and up to 400 iterations of the Nelder-Mead algorithm successfully converged all values of $\phi$ to within $||\phi|| \leq  0.02$. This suggests the controller can successfully align any detuning, as long as the each ring is able to tune over an FSR. Fig. \ref{fig:simulation_5ring_randominitial} plots the convergence of 10 of the random  detunings that initially have $||\phi|| \geq \pi$.  Appendix \ref{sec:example convergence} shows how the passband response is progressively recovered for one of the initial conditions.

\begin{figure}[t!]
\centering
\includegraphics[width = \columnwidth]{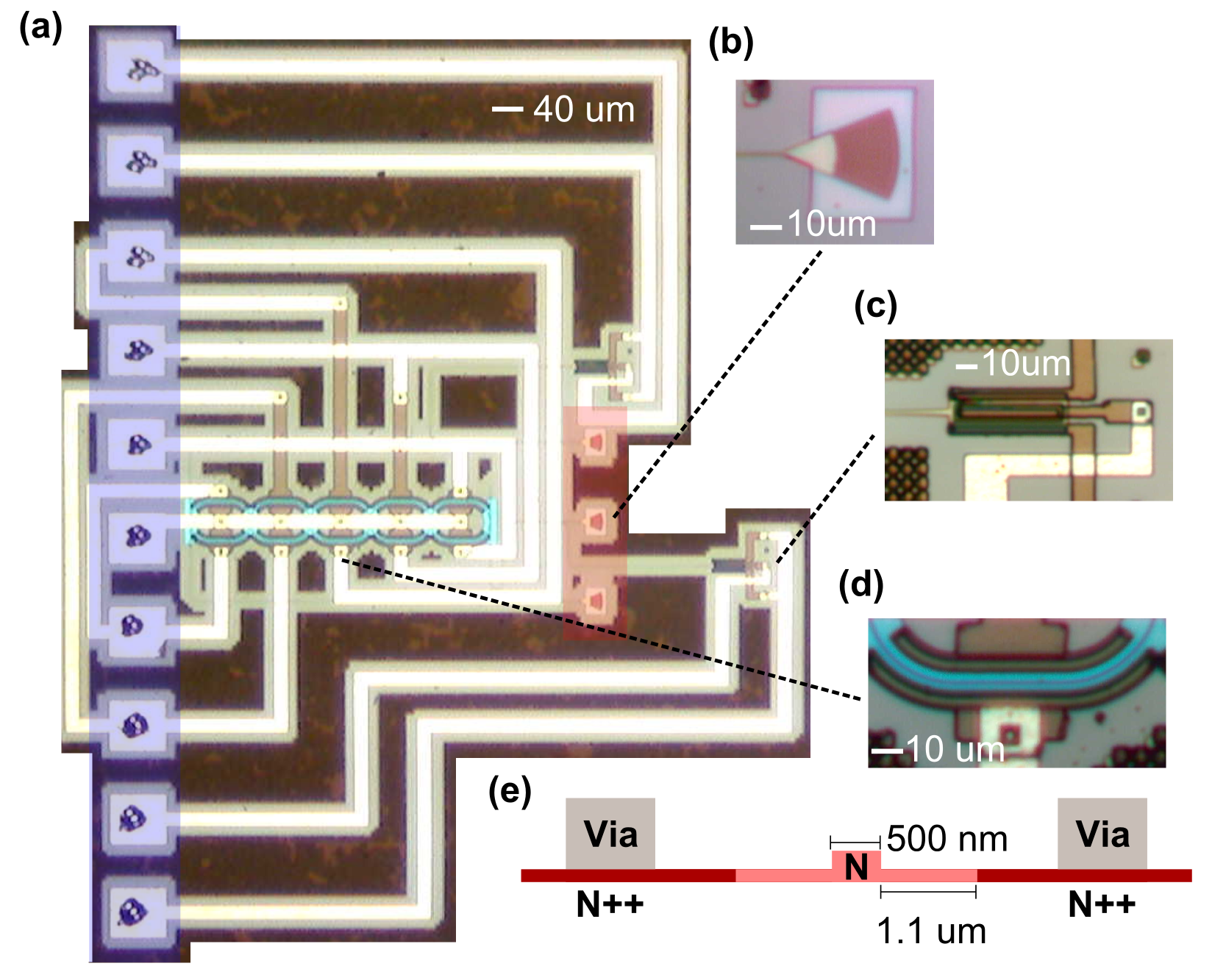}
\caption{(a) Microscope image of the fabricated device. Contact pads for electrical probing are highlighted in blue on the left side. Grating couplers at a 127 $\mu$m pitch for coupling to a fiber array are highlighted in red on the right side. A close-up view of (b) the grating coupler, (c) germanium photodiode, and (d) integrated heater. (e) A cross-section schematic of the integrated heater.}
\label{fig:fabricated_ring_system}
\end{figure}

The 5-ring filter was fabricated through the A*STAR IME baseline active silicon photonics process on 8'' SOI wafers \cite{CMCIME, Liow2010}.  A microscope image of this device is shown in Fig. \ref{fig:fabricated_ring_system}(a). Input and output optical signals are coupled to an array of standard single-mode fibers (SMF-28) polished and tilted at an $8^\mathrm{o}$ angle through grating couplers spaced at a pitch of 127 $\mathrm{\mu}$m. The ring waveguides were as described previously.  The microrings were rounded rectangles, with corners consisting of circular $90^\mathrm{o}$ bends with a radius of $30$ $\mathrm{\mu m}$. Opposite sides forming directional couplers had lengths of 2.5 $\mathrm{\mu m}$, and the remaining sides forming the resistive heaters had a length of 35 $\mathrm{\mu m}$. The coupling gaps were designed to be 230 nm, 380 nm, 430 nm, 430 nm, 380 nm, and 230 nm, for $\kappa = [\kappa_0,\dots,\kappa_5]$.  Doped Si resistive heaters were formed in the microring.  A heater consisted of an N doped region in the waveguide sandwiched by N++ doped contacts 1.1 $\mathrm{\mu}$m away from the sides of the waveguide.  The N++ regions connected to metal vias and contacts. The doping covered the straight section and half of the ring bends. Each microring had an independent heater.  At the drop port,  25\% of the power was tapped using a directional coupler and measured by an integrated  germanium photodetector. The thru port also had a 25\% power tap which was not used in the experiments to follow.  The grating coupler, germanium photodectector, and resistive heaters are highlighted in Figs. \ref{fig:fabricated_ring_system}(b), (c), (d), respectively. The heater cross-section is in Fig. \ref{fig:fabricated_ring_system}(e).

\subsection{Setup and Procedure}
\label{sec:setup and procedure}
\begin{figure}[t!]
\centering
\includegraphics[width = \columnwidth]{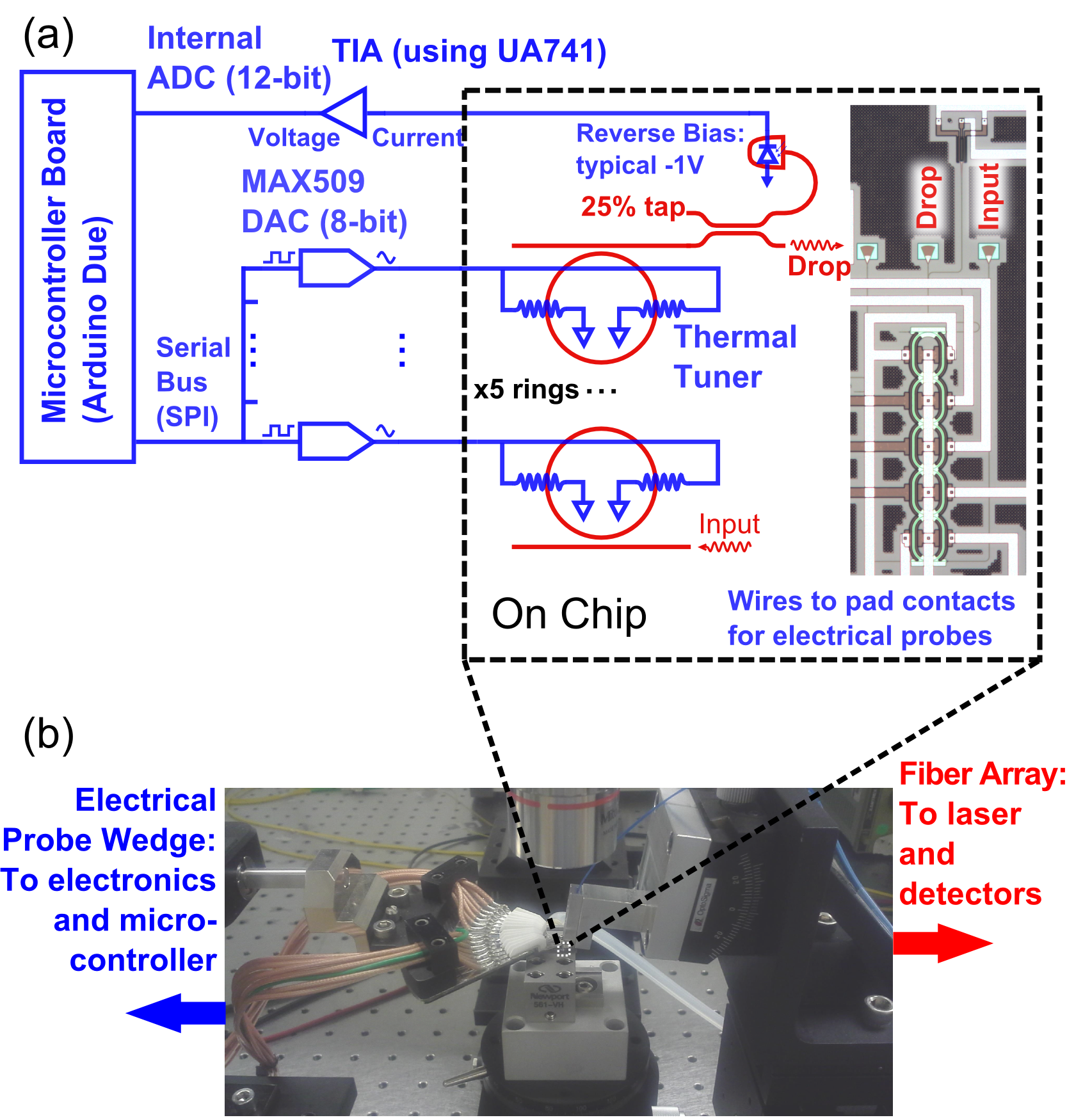}
\caption{(a) Schematic diagram of the electro-optical feedback system. (b) Photograph of the probing setup.}
\label{fig:experimental_setup}
\end{figure}

An Arduino Due \cite{Arduino} microcontroller board controlled by a  computer  executed the controller algorithm. Here, for simplicity, the processing for the controller algorithm was performed on the computer, and the microcontroller board mainly served as an interface for the electronics to facilitate debugging and monitoring.  $\phi_{\textrm{tuning}}$ was changed by applying voltages across the heaters, and the drop port power was monitored by reading the photocurrent. To apply voltages to the 5 heaters, we used digital-to-analog converter (DAC) chips (MAX509) which were controlled through a Serial Peripheral Interface (SPI) bus by the microcontroller. The DACs provided a voltage ranging from 0 to 3 V with a resolution of 256 steps. To read the photocurrent, we implemented a basic transimpedance amplifier circuit using an op-amp (UA741). A simplified schematic of the feedback system hardware is shown in Fig. \ref{fig:experimental_setup}(a). The electronics were connected to the device using an electrical multi-contact wedge. The probing of the device is pictured in  Fig. \ref{fig:experimental_setup}(b).   The components used in this demonstration were sourced off-the-shelf, though custom integrated circuitry could be designed to be more densely integrated with the device if needed.  

During the resonance alignment, input laser light supplied by a  tunable laser source was set to the desired center wavelength for the optical filter. A reverse bias was applied to the photodiode, ranging from 0 to -1.5V.  5 to 10 full cycles of the coordinate descent were applied, followed by 30 to 50 iterations of the Nelder-Mead simplex algorithm. An iteration of the coordinate descent algorithm was implemented by sweeping the voltage across the discrete steps while monitoring the photocurrent reading, and choosing the tuning corresponding to the maximum reading. The Nelder-Mead algorithm used the SciPy \cite{scipy} implementation in the {\tt scipy.optimize} library. Although the algorithm outputs floating point numbers, the values were rounded to the nearest integer, and saturated at 0 and 255 to make the output compatible with the DAC. 

\subsection{Results}
\begin{figure}[t!]
\centering
\includegraphics[width = 0.4\columnwidth]{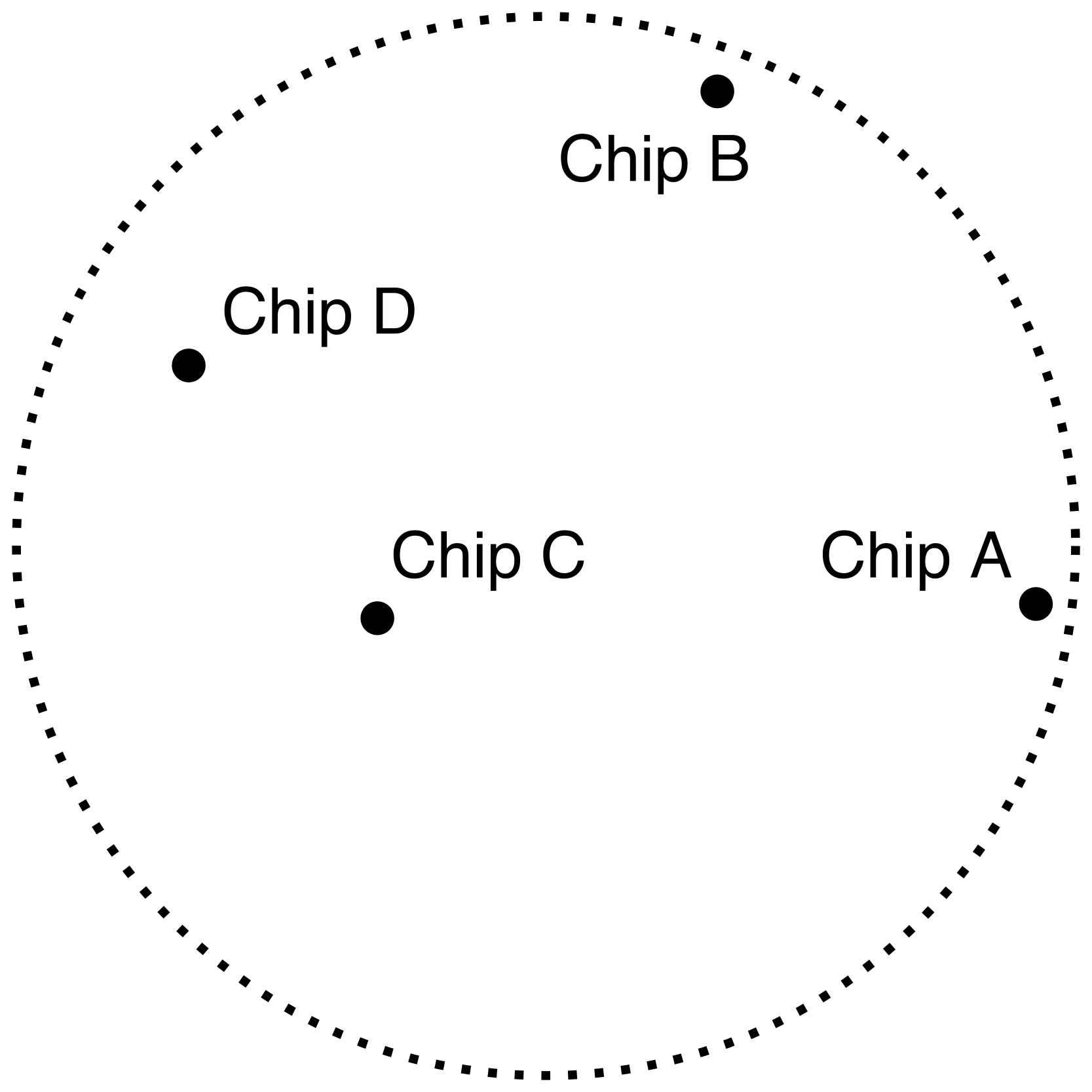}
\caption{Wafer map of the devices tested. The wafer was 8'' in diameter.}
\label{fig:wafer_map}
\end{figure}

We tested the automatic resonance alignment with nominally identical 5-ring filters taken from 4 dies across the 8'' wafer at the locations shown in Fig. \ref{fig:wafer_map}.  The measured 5-ring devices were tunable over an FSR within the 3 V tuning range.  The spectra shown in Figs. \ref{fig:results_5ring_randominitial},  \ref{fig:results_5ring_reliability}, \ref{fig:results_5ring_wavelength} were normalized against the maximum thru port power to remove losses from the grating couplers. In the discussion to follow, we describe the ability of the system to recover the passband spectrum from arbitrary initial spectra, the repeatability of the alignment, and the wavelength tunability of the alignment.

\subsubsection{Varying the Starting Spectrum}
\begin{figure}[t!]
\centering
\includegraphics[width = \columnwidth]{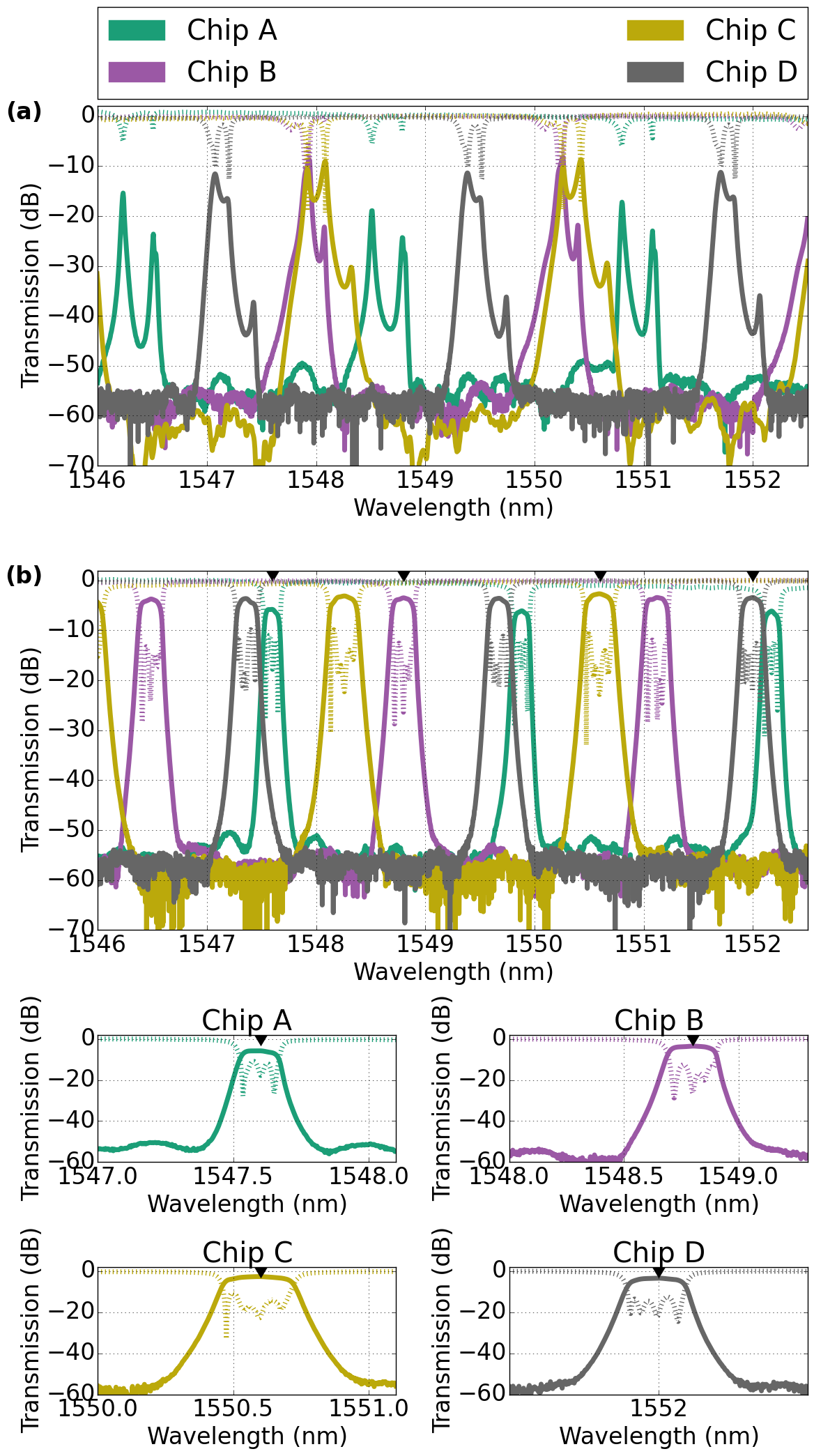}
\caption{(a) The measured initial transmission spectra of the devices from 4 chips across the wafer.  The detuning varied widely across the wafer. (b) The top part shows the measured resonance aligned transmission spectra across a few FSRs.  For clarity, each chip was set to a different center wavelength, which is indicated by a black marker.  The bottom part shows magnified views of a passband for each chip. }
\label{fig:results_5ring_randominitial}
\end{figure}

Fig. \ref{fig:results_5ring_randominitial} shows the automatic resonance alignment of devices from the 4 dies corresponding to the locations on Fig. \ref{fig:wafer_map}.  The initial spectra of the as-manufactured devices are shown in Fig. \ref{fig:results_5ring_randominitial}(a), and they exhibited significant variability, sharp ripples and were centred at various wavelengths.  The insertion loss of the filter was ranged between 18.92 dB to 8.98 dB, and the bandwidth of the filter was not well defined.  Fig. \ref{fig:results_5ring_randominitial}(b) shows the spectral response after alignment for the reference wavelengths indicated by the black markers.  Table \ref{tab:randominitial} summarizes the final filter characteristics after the resonance alignment.  Flat-top passbands were recovered in all instances, and the insertion loss was reduced to $3.67 \pm 1.8$ dB.  The FSR of the filters ranged between 2.29 nm and 2.34 nm.  $\Delta \lambda_{\mathrm{1 dB}}$ was about $0.15 \pm 0.3$ nm.  Within $\Delta \lambda_{\mathrm{1 dB}}$, the ripples in the drop port transmission spectra were less than 0.2 dB.  The variation in the final filter response was most likely due to variations in the coupling coefficients, because  a narrower bandwidth was associated with a higher insertion loss, as expected from weaker inter-ring coupling \cite{PoonJOSAB2004}.  In Si photonic platforms, the coupling coefficients of directional couplers are  highly sensitive to fabrication variation \cite{mikkelsen2014dimensional}. Despite the large variation in the initial detuning and coupling in the devices, a passband response could be recovered automatically by this system.

\begin{table}
\centering
\caption{Final filter properties for devices on different dies corresponding to Fig. \ref{fig:results_5ring_randominitial}(b)}
\label{tab:randominitial}
    \begin{tabular}{l p{0.8cm}  p{0.8cm}  p{0.8cm} p{0.6cm}  p{0.6cm}  p{0.6cm}}
    \hline
& Reference Wavelength (nm) & Center Wavelength (nm) & Insertion Loss (dB) &  FSR (nm) & $\Delta \lambda_{\mathrm{1 dB}}$ (nm) & $\Delta \lambda_{\mathrm{3 dB}}$ (nm)\\ \hline
Chip A &1547.60 &1547.60  &-5.47 & 2.29&0.11 &0.14  \\
Chip B &1548.80 &1548.80&-2.80  &2.32 &0.16 &0.19  \\ 
Chip C &1550.60 &1550.59 &-1.86 &2.34 &0.18 &0.21  \\ 
Chip D &1552.00 &1552.00  &-3.05 &2.33 &0.14 &0.19  \\ 
    \hline
    \end{tabular}
\end{table}

\subsubsection{Repeatability}

\begin{figure}[t!]
\centering
\includegraphics[width = \columnwidth]{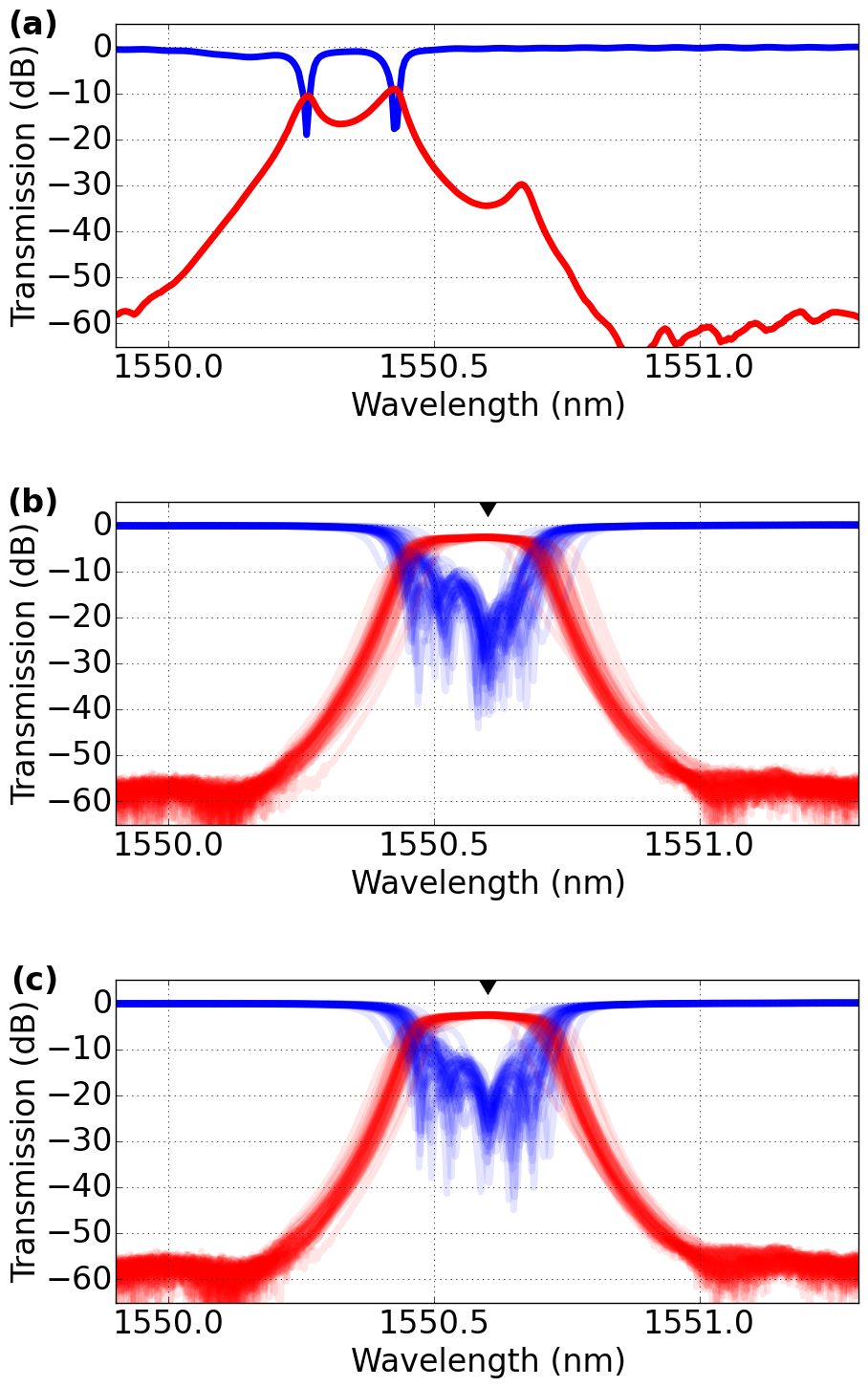}
\caption{(a) The measured initial thru and drop  spectra of the 5-ring filter in Chip C. The algorithm is applied to the same device for 50 trials.  The measured thru and drop port spectra after (b)  the coordinate descent stage and (c) Nelder-Mead simplex stage for the 50 trials. The statistics of the recovered passband are summarized in Table \ref{tab:reliability stage1}.} %chip C is chip 25
\label{fig:results_5ring_reliability}
\end{figure}

To verify the repeatability of the resonance alignment, we ran 50 consecutive trials on the device from Chip C.  Figures  \ref{fig:results_5ring_reliability}(a), (b), and (c) respectively show the spectra before the tuning, after the coordinate descent stage for the  50 trials, and after the Nelder-Mead simplex method stage for the 50 trials. In Figs.  \ref{fig:results_5ring_reliability}(b) and (c), datasets for the trials have been superimposed to show the consistency between the resulting drop port spectra. Variations existed between the trials due to noise from the photocurrent, electronics, and discretization.  As expected from the simulations, the results after the Nelder-Mead stage had less variance than that from the coordinate descent stage, as seen in Figs. \ref{fig:results_5ring_reliability}(b) and (c) and  summarized in Table \ref{tab:reliability stage1}. For these measurements, the reference wavelength was set to 1550.6 nm. The centre wavelength of the filter responses after coordinate descent was 1550.58 nm with a standard deviation of 0.024 nm, and the centre wavelength after the Nelder-Mead stage was 1550.50 nm with a standard deviation of 0.016 nm.  The insertion loss after the coordinate descent stage was 2.64 dB had a standard deviation of 0.13 dB, which was reduced to 2.58 dB with a standard deviation of 0.11 dB after the Nelder-Mead stage. The 3 dB bandwidth of the filter was also reduced from 0.247 nm with a standard deviation of 0.096 nm to 0.24 nm with a standard deviation of 0.02 nm between the coordinate descent and Nelder-Mead stages. The results show that the Nelder-Mead optimization was critical in providing a more repeatable alignment, especially in the bandwidth of the filter.

\begin{table}
\begin{center}
\caption{Variation in the filter properties for a single device over 50 trials corresponding to Fig. \ref{fig:results_5ring_reliability}}\label{tab:reliability stage1}
    \begin{tabular}{ l  l  l  l  p{0.8cm}}
    \hline
     & Mean & Max & Min & Std. Dev.\\ \hline
    \textbf{After Coordinate Descent} &&&\\
    Center Wavelength (nm) & 1550.58 & 1550.62 & 1550.50 & 0.024 \\ 
	Insertion Loss (dB) & 2.64 & 3.07 & 2.46 & 0.13 \\
	3 dB Bandwidth (nm) & 0.247 & 0.909 & 0.175 & 0.096 \\ 
	\\
	\textbf{After Nelder-Mead}&&&\\
	Center Wavelength (nm) & 1550.60 & 1550.62 & 1550.52 & 0.016 \\ 
	Insertion Loss (dB) & 2.58 & 3.03 & 2.38 & 0.11 \\ 
	3 dB Bandwidth (nm) & 0.240 & 0.26 & 0.16 & 0.02 \\ 
    \hline
    \end{tabular}
\end{center}
\end{table}

\subsubsection{Wavelength Tuning}

\begin{figure}[t!]
\centering
\includegraphics[width = \columnwidth]{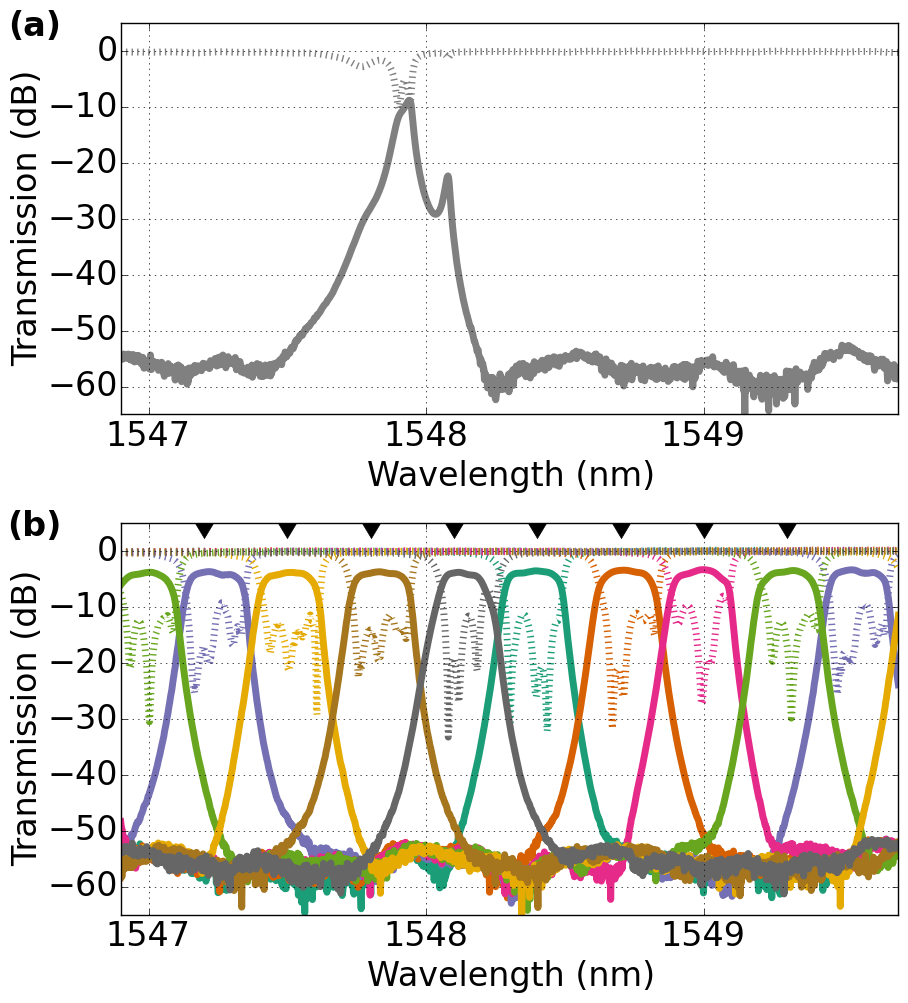}
\caption{(a) The measured initial thru and drop spectra of the 5-ring filter in Chip B. Thru spectra are in dotted lines, and drop spectra are in solid lines. (b) The device is tuned to various center wavelengths, as indicated by the black triangles. The passband characteristics are summarized in Table \ref{tab:wavelength}.} %Chip B is Chip 3
\label{fig:results_5ring_wavelength}
\end{figure}

Lastly, to demonstrate that the alignment system can be used for wavelength tuning, we varied the reference wavelength to show that a flat-top passband could be recovered over the FSR.  The device was designed to be tunable over the FSR, so the filter passband could be set to an arbitrary wavelength. Using the device on Chip B, we stepped the centre wavelength in increments of 0.3 nm over a 2.1 nm range, returning to the initial spectrum  between successive alignments. The initial spectrum of the device is shown in Fig. \ref{fig:results_5ring_wavelength}(a), and the wavelength tunability is shown in Fig. \ref{fig:results_5ring_wavelength}(b).  The characteristics of the filter are summarized in Table \ref{tab:wavelength}.  Within the tuning range, the centre wavelength of the filter was accurate to within 0.02 nm; the insertion loss varied within 0.28 dB; and   $\Delta \lambda_{\mathrm{1 dB}}$ and $\Delta \lambda_{\mathrm{3 dB}}$ varied within 0.04 nm.

\begin{table}
\begin{center}
\caption{Filter properties as the reference wavelength is tuned corresponding to Fig. \ref{fig:results_5ring_wavelength}}
\label{tab:wavelength}
    \begin{tabular}{ p{1.5cm}  p{1.5cm} p{0.8cm}  p{0.6cm} p{0.8cm}  p{0.8cm}}
    \hline
Reference Wavelength (nm) & Center Wavelength (nm) & Insertion Loss (dB) & FSR (nm) & $\Delta \lambda_{\mathrm{1 dB}}$ (nm) &$\Delta \lambda_{\mathrm{3 dB}}$  (nm)\\ \hline
1547.20 &1547.22 &-3.33 &2.32 &0.16 &0.20 \\
1547.50 &1547.51 &-3.51 &2.31 &0.15 &0.20 \\
1547.80 &1547.83 &-3.47 &2.32 &0.16 &0.20 \\
1548.10 &1548.12 &-3.49 &2.32 &0.12 &0.16 \\
1548.40 &1548.39 &-3.55 &2.33 &0.16 &0.20 \\
1548.70 &1548.71 &-3.27 &2.32 &0.14 &0.19 \\
1549.00 &1549.00 &-3.42 &2.33 &0.13 &0.20 \\
1549.30 &1549.32 &-3.38 &2.31 &0.14 &0.20  \\ \hline
Mean & N/A & -3.42 & 2.32 &0.15 & 0.19 \\
Min & N/A & -3.27 &2.31 & 0.12 & 0.16 \\
 Max & N/A & -3.55 &2.33& 0.16 & 0.20\\
Std. Dev & N/A & -0.09 & 0.01 & 0.01 & 0.01\\
    \hline
    \end{tabular}
\end{center}
\end{table}

\section{Discussion}\label{sect:discuss}

%consistency with simulation results
The results show that the implemented automatic resonance alignment system behaves in agreement with the simulations, even in the presence of non-idealities such as noise and discretization. The simulations showed the recovery of the passband as $\phi$ converged to $ 0 \Mod{2\pi}$ from a random initial detuning, which was confirmed in practice with Fig. \ref{fig:results_5ring_randominitial}. The alignment was repeatable for the same reference wavelength setting and for different wavelengths (Figs. \ref{fig:results_5ring_reliability} and \ref{fig:results_5ring_wavelength}). Because the devices tested were arbitrarily sampled from across the wafer, and even at the edge of the wafer, this lends confidence that the automatic alignment is effective on the fabrication variations found in typical foundry fabricated devices.  However, since the coupling coefficients and losses were not tunable in the devices presented here, the final spectra for different starting conditions exhibited variations in the bandwidth and insertion loss.  The variations were more obvious in the thru port spectra. 

%non-idealities
In the experiments, there was variability in the recovered spectra even for the same starting conditions, as evidenced by the consecutive trials in Fig. \ref{fig:results_5ring_reliability} and Table \ref{tab:reliability stage1}. The Nelder-Mead stage of the algorithm was responsible for bringing the result close to resonance, and involved increasingly fine adjustments to the phase. The implementation of the Nelder-Mead algorithm assumed a continuous search domain, but with discretization and noise,  the resolution of the adjustments was limited in practice.   Therefore, the variation in the recovered spectra may be reduced by using a more  sophisticated algorithm which takes into account the discretization and noise or by using more sensitive and higher resolution electronics that can make the alignment more accurate. 

Although in this work we have demonstrated this system for a 5-ring filter, the theory is independent of the number of rings. An even higher number of rings can be aligned in principle, if the drop port power signal is strong enough to be monitored by the photodiode. Compared with the direct monitoring of the power in each ring, the advantage of our method is a reduction of the required number of monitors. Using our proposed optimization method, the automatic alignment of high-order microring filters can be achieved with relatively simple electronics  (only one photodiode has to be monitored) and more flexible design of the rings not constrained by placement of monitors. The resonance alignment was achieved by treating the propagation phase as independent variables, and this concept of phase tuning through feedback can be extended to other types of interference-based photonic devices and circuits, such as Vernier filters, networks of Mach-Zehnder interferometer filters, and optical switch fabrics \cite{LeeJLT2014, Sherwood-DrozOE2008, chu2004compact, horst2013cascaded}. 

Finally, we note some limitations of this work and comment on pathways for future improvement. In this work, for flexibility and ease of programming, the resonance alignment algorithm was executed on a computer, which sent the results to the microcontroller board. The microcontroller board can be programmed to execute the algorithm to eliminate the use of a computer to simplify the hardware required by the system. The algorithm can be modified to accommodate filters that have stronger ripples in the passband, for example, ideal Chebyshev or elliptic filters, since the direct optimization of $|T_D(\phi)|^2$ may end up in a local maximum from the ripples. Currently, losses from the power tap at the drop port contribute to the insertion loss of the device, and a non-invasive monitor, such as the contactless photonic probe proposed in \cite{MorichettiJSTQE2014, GrillandaOPT2014}, can remove this insertion loss entirely without affecting the method. Variability in the coupling resulted in variability in bandwidths and insertion losses; this could be mitigated by using variational tolerant couplers \cite{mikkelsen2014dimensional}, multimode-interference couplers, or tunable couplers \cite{orlandi2013tunable,sacher2013coupling, sacher2014binary}. Finally, by using more sophisticated controller designs \cite{sastry2011adaptive}, the ideas in this work can be extended to perform wavelength tracking and stabilization against temperature and environmental influences in high-order microring filters.  

\section{Conclusion}

In summary, we have demonstrated a feedback system which automatically aligns the resonances of high-order microring filters and can potentially be scaled to high-volume production.  Automatic tuning is critical for making use of complex Si photonic devices and circuits because of the high sensitivity to fabrication variation in high optical confinement, sub-micron, Si waveguides.  Our method uses a combination of coordinate descent and Nelder-Mead simplex optimization algorithms to converge to a phase compensation that maximizes the drop port power at a  reference wavelength.  We demonstrated the automatic resonance alignment using Si 5-ring filters that were fabricated in a foundry process.  The results show that the method is robust, and it can be applied to a variety of starting spectra found across the wafer and arbitrary wavelengths within the FSR.  This work can be extended to stabilize the operation of the optimized filter.  Also, the strategy of propagation phase tuning through feedback can be applied to other types of interference-based Si photonic devices and circuits.

\section*{Acknowledgments}
The authors are grateful for the financial support of the Natural Sciences and Engineering Research Council of Canada and the Canada Research Chairs program.   The device fabrication using the A*STAR IME baseline silicon photonic process was supported by CMC Microsystems.  The assistance of Dan Deptuck and Jessica Zhang of CMC Microsystems is gratefully acknowledged.

%%%%%%%%%%%%%%
%%%% APPENDIX %%%%%
%%%%%%%%%%%%%%
\appendices
\section{Example Convergence of a 5-Ring Filter}
\label{sec:example convergence}

\begin{figure*}%[t!] %use * environment to make it span across 2 pages?
\centering
\includegraphics[width = \textwidth]{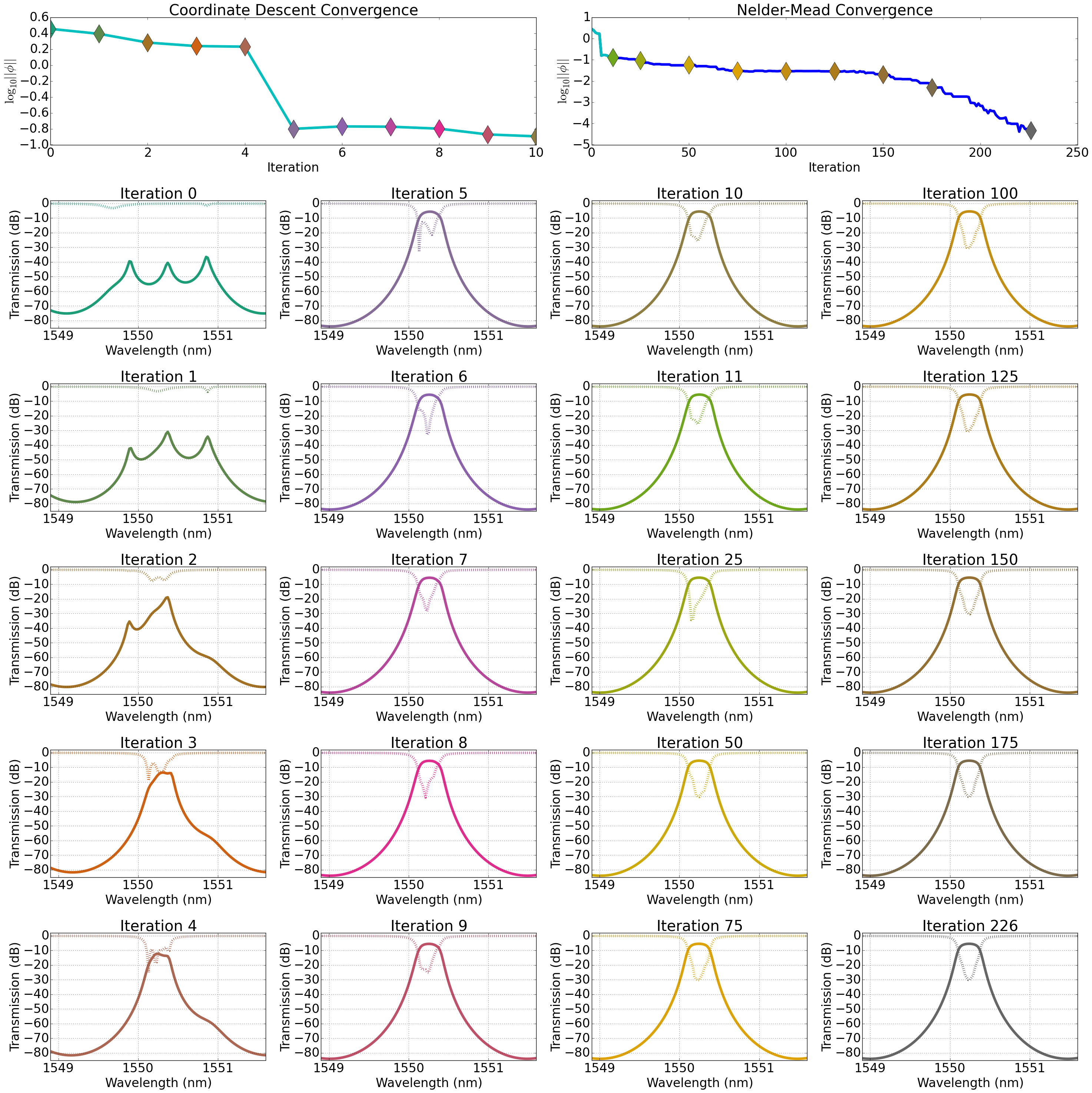}
\caption{Visualizations of the convergence (top row) and transmission spectra for intermediate iterations from a simulation of the resonance alignment, with the initial detuning of $\phi = [-1.43, 1.56, -0.83, 0.34, 1.7]$.}
\label{fig:simulation_5ring_stepped}
\end{figure*}

In this appendix, we examine the spectrum of the 5-ring filter during the intermediate steps of the resonance alignment. The model of the 5-ring filter used the parameters described in Section \ref{sect:device design}.  We start from the initial detuning of $\phi = [-1.43, 1.56, -0.83, 0.34, 1.7]$ and proceed with 10 iterations of coordinate descent and at most 400 iterations of the Nelder-Mead algorithm. The Nelder-Mead algorithm converged after 216 iterations with 346 function evaluations. In Fig. \ref{fig:simulation_5ring_stepped}, we plot the convergence of the coordinate descent and the Nelder-Mead portions at the top. At the bottom, we show the  progression of the spectrum from the initial state at Iteration 0 to the end of the coordinate descent algorithm at Iteration 10, and more sparsely to the end of the optimization at Iteration 226. As expected, the large jumps in convergence in the coordinate descent within the first 5 iterations correspond to the largest changes to the spectrum overall. After the coordinate descent, changes to the drop port are small, and the convergence toward $||\phi|| = 0$ is only appreciable from changes in the thru port spectrum.

\bibliographystyle{IEEEtran}
%\bibliography{../bibtex/JQE2015-Introduction,../bibtex/theory,../bibtex/JQE2015-Experiment} %no whitespace in arguments!

\begin{IEEEbiographynophoto}{Jason C. C. Mak} received the B.A.Sc. degree in engineering science (physics option) from the University of Toronto, Toronto, ON, Canada, in 2013.

He is a Ph.D. candidate in the Department of Electrical and Computer Engineering at the University of Toronto, where he holds the NSERC Canada Graduate Scholarship at the masters level.  He held the Ontario Graduate Scholarship between 2013 and 2014.  In 2011, he was an intern for the Hardware Qualification Signal Integrity group in AMD Markham.

Mr. Mak's research interests include microring devices, device optimization and control systems for photonic devices.
\end{IEEEbiographynophoto}

\begin{IEEEbiographynophoto}{Wesley D. Sacher} received the B.A.Sc. degree in engineering science (electrical
option) from the University of Toronto, Toronto, ON, Canada, in 2009.

He is  a Ph.D. candidate in the Department of Electrical and Computer Engineering at the University of Toronto.  He held  NSERC Canada Graduate Scholarships at the masters and doctoral levels between 2010 and 2014.  In 2011, he was an intern in the Silicon Integrated Nanophotonics group at IBM T. J. Watson Research Center.  He has been a part-time researcher for IBM since 2013. 

Mr. Sacher's research interests include integrated photonics, microring modulators, polarization management, lasers, and nanofabrication.
\end{IEEEbiographynophoto}

\begin{IEEEbiographynophoto}{Tianyue Xue} is an undergraduate student in the engineering science (physics option) program at the University of Toronto, Toronto, ON, Canada.  He expects to graduate in 2016. Mr. Xue is interested in electronics and applications of embedded systems.

\end{IEEEbiographynophoto}

\begin{IEEEbiographynophoto}{Jared C. Mikkelsen} received the B.A.Sc. degree in engineering science (physics
option) from the University of Toronto, Toronto, ON, Canada, in 2011.

He is  a Ph.D. candidate in the Department of Electrical and Computer Engineering at the University of Toronto, where he holds an NSERC Postgraduate Graduate Scholarship at the doctoral level. In 2015, he was an intern in the silicon photonics group at Finisar.

Mr. Mikkelsen's research interests include silicon photonics and microring modulators.

\end{IEEEbiographynophoto}

\begin{IEEEbiographynophoto}{Zheng Yong} received the B.Eng. degree in optical engineering from Zhejiang University, Hangzhou, Zhejiang, China in 2013.

He is  a Ph.D. candidate in the Department of Electrical and Computer Engineering at the University of Toronto.  His research interests include high-speed modulators and photodetectors, optoelectronic design for silicon photonics, and nanofabrication.
\end{IEEEbiographynophoto}

\begin{IEEEbiographynophoto}{Joyce K. S. Poon}(S'01, M'07) received the B.A.Sc. degree in engineering science (physics
option) from the University of Toronto, Toronto, ON, Canada, in 2002, the M.S. and Ph.D. degrees in electrical engineering from the California Institute of Technology, Pasadena, in 2003 and 2007, respectively.  

She is presently an Associate Professor in the Department of Electrical and Computer Engineering at the University of Toronto, where she holds the Canada Research Chair in Integrated Photonic Devices.  She and her team conduct theoretical and experimental research in micro- and nano-scale integrated photonics implemented in  silicon-on-insulator (SOI), indium phosphide on SOI, silicon nitride on SOI, indium phosphide, and correlated electron materials.

\end{IEEEbiographynophoto}

\end{document}